\documentclass[journal]{IEEEtran}
\usepackage{graphicx}
\usepackage{amsmath,amssymb,amsfonts}
\usepackage{algorithm}
\usepackage{algorithmic}
\usepackage{psfrag}
\usepackage{subfigure}
\usepackage{color}
\usepackage{soul}
\usepackage{xcolor}
\usepackage[hyphens]{url}
\usepackage[bookmarks=false]{hyperref}
\usepackage[hyphenbreaks]{breakurl}
\usepackage{lipsum}
\usepackage{pbox}
\usepackage{multicol}
\usepackage{xtab}
\usepackage{booktabs}
\usepackage{array}
\usepackage{tikz}
\newcommand*\circled[1]{\tikz[baseline=(char.base)]{
            \node[shape=circle,fill,inner sep=0.5pt] (char) {\textcolor{white}{#1}};}}

\makeatletter
\let\mcnewpage\newpage
\newcommand{\changenewpage}{%
  \renewcommand\newpage{%
    \if@firstcolumn
      \hrule width\linewidth height0pt
      \columnbreak
    \else
      \mcnewpage
    \fi
}}
\makeatother

\graphicspath{{Figures/}}
\ifCLASSINFOpdf
\else
\fi

\def\rot#1{\rotatebox{90}{#1}}

\ifCLASSINFOpdf
\else
\fi
\begin{document}

\title{L-Platooning: A Protocol for Managing a Long Platoon with DSRC}

\author{TBD}
\author{Myounggyu~Won\\
Department of Computer Science\\University of Memphis, Memphis, TN, United States\\

mwon@memphis.edu}%

\markboth{IEEE Transactions on Intelligent Transportation Systems}%
{Shell \MakeLowercase{\textit{et al.}}: Bare Demo of IEEEtran.cls for Journals}

\maketitle

\begin{abstract}
Vehicle platooning is an automated driving technology that enables a group of vehicles to travel very closely together as a single unit to improve fuel efficiency and driving safety. These advantages of platooning attract huge interests from academia and industry, especially logistics companies that can utilize the platooning technology for their heavy-duty trucks due to the huge cost savings. In this paper, we demonstrate that existing platooning solutions, however, fail to support formation of a `long' platoon consisting of many vehicles especially long-body heavy-duty trucks due to the limited range of vehicle-to-vehicle communication such as DSRC and device-to-device communication for C-V2X. To address this problem, we propose \emph{L-Platooning}, the first platooning protocol that enables seamless, reliable, and rapid formation of a long platoon. We introduce a novel concept called \emph{Virtual Leader} that refers to a vehicle that acts as a platoon leader to extend the coverage of the original platoon leader. A virtual leader election algorithm is developed to effectively designate a virtual leader based on the novel metric called the \emph{Virtual Leader Quality Index (VLQI)} which quantifies the effectiveness of a vehicle serving as a platoon leader. We also develop mechanisms for \emph{L-Platooning} to support the vehicle join and leave maneuvers specifically for a long platoon. Through extensive simulations, we demonstrate that \emph{L-Platooning} enables vehicles to form a long platoon effectively by allowing them to maintain the desired inter-vehicle distance accurately. We also show that \emph{L-Platooning} handles seamlessly the vehicle join and leave maneuvers for a long platoon.
\end{abstract}

\begin{IEEEkeywords}
Vehicle platooning, vehicle to vehicle communication (V2V), autonomous driving, truck platooning
\end{IEEEkeywords}

%
\IEEEpeerreviewmaketitle

\section{Introduction}
\label{sec:introduction}


More than 71\% of freight in U.S. are transported by trucks~\cite{ATA}. According to the U.S. Bureau of Transportation Statistics~\cite{BTS}, the total weight of goods transported by trucks is 18.6 billion tons in 2018, and the value of which is \$18.9 trillion. It is estimated that there are over 15 million trucks registered for business purposes, 3.5 million truck drivers, and over 1.2 million freight companies in U.S. only~\cite{TruckStat}. Due to the huge number of trucks on the road, the fuel consumption of trucks is about 12.8\% of the total fuel consumed in the country~\cite{TruckStat}. According to American Transportation Research Institute (ATRI), trucks in U.S. consume 53.9 billion gallons of fuel in a year leading to the annual operation cost of over \$21.4 billion~\cite{ATRI}. Besides the financial burden, transportation is the biggest source of carbon dioxide (CO2) emissions, producing 1.9 billion tons of CO2 every year~\cite{Yale}. U.S. Environmental Protection Agency (EPA) reported that 23\% of these transportation-related CO2 emissions are emitted by medium and heavy duty trucks, accounting for about 6\% of total CO2 emissions in U.S.~\cite{EPA}.

A new driving technology called the vehicle platooning has been developed to address these financial and environmental problems~\cite{hall2005vehicle}. The vehicle platooning basically allows vehicles to drive as a single unit by automatically maintaining small inter-vehicle distance. It is powered by systematic coordination of different technologies including vehicle control \cite{rajamani2000demonstration}\cite{ploeg2011design}\cite{di2014distributed} \cite{santini2015consensus}, sensing, and vehicle-to-vehicle (V2V) communication technology such as IEEE 802.11p (DSRC)~\cite{vukadinovic20183gpp} and 3GPP/LTE-based C-V2X~\cite{3gpptech}. Specifically, the vehicle control system is designed to adjust the acceleration of the vehicle such that a constant inter-vehicle distance to the preceding vehicle is maintained; The V2V communication and vehicular sensor systems support the operation of the vehicle control system by providing the kinematic information of the ego vehicle as well as surrounding vehicles via wireless communication.

Development of the vehicle platooning dates back to 1970~\cite{dadras2015vehicular} and started to receive significant attention from academia and industry in 1990s when the California
Partners for Advanced Transit and Highways (PATH) program was initiated~\cite{shladover2007path}. Since then, numerous technologies for enhancing platooning have been investigated all over the world~\cite{coelingh2012all}. Researchers demonstrated that vehicle platooning has huge potential to address numerous transportation problems~\cite{van2006impact}. For example, platooning can be used to reduce traffic congestion and increase the road capacity because in platooning, vehicles drive together with very small inter-vehicle distances. In particular, the fuel efficiency is significantly enhanced as the aerodynamic drag is substantially reduced because of the extremely small gap between vehicles~{\cite{lyamin2016study}}. Such high fuel efficiency leads to reduced CO2 emissions. Furthermore, since a vehicle in a platoon autonomously follows the preceding vehicle, driving can be more comfortable and safer~{\cite{axelsson2016safety}}.

\begin{figure}
\centering
\includegraphics[width=.9\columnwidth]{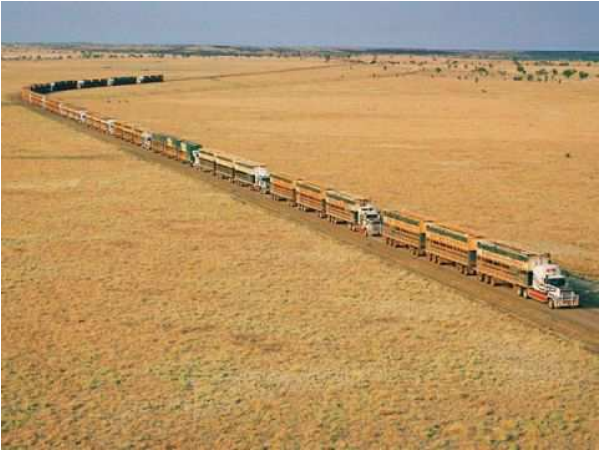}
\caption {An example of a ``road train'' that consists of a large number of trailer trucks drive as a single unit~\cite{RoadTrain}.}
\label{fig:long_platoon}
\end{figure}

In this paper, we demonstrate that existing vehicle platooning solutions, however, fail to support formation of a `long' platoon consisting of many vehicles especially long-body heavy-duty trucks (\emph{e.g.,} Fig.~{\ref{fig:long_platoon}}) due to the limited range of vehicle-to-vehicle communication such as DSRC and device-to-device mode for C-V2X. More precisely, we show that considering the minimum allowable length limit for a trailer of 14.63m~\cite{TruckSize}, the standard cabin length, and a typical inter-vehicle distance between 10m and 20m, only less than 10 trailer trucks can form a platoon reliably due to the limited range. This is a significant disadvantage for freight companies that run a huge number of heavy-duty trucks. Specifically, in order for the vehicle control system to maintain the constant inter-vehicle distance accurately, it is crucial for the platoon members to receive the messages containing the kinematic information such as the speed and acceleration of other platoon members via V2V reliably. However, due to the limited range, the platoon members that are out of the range of the platoon leader may fail to receive the messages, which can affect the performance of maintaining the inter-vehicle distance reliably.

In this paper, we present \emph{L-Platooning}, a platooning protocol to support effective and seamless formation of such `long' platoons. While the importance of utilizing long platoons have been mentioned in the literature~{\cite{davis2018dynamics}\cite{robertson2019experimental}\cite{gao2019scalable}\cite{goli2020merging}}, this is the first working protocol that is designed to support formation of a long platoon under dynamic network conditions where the link quality between two vehicles change over time, specifically addressing the problem of the limited range of V2V communication. The proposed protocol allows vehicles to coordinate autonomously and form a long platoon seamlessly. In \emph{L-Platooning}, a novel concept called \emph{Virtual Leader} is introduced which refers to a platoon member that acts as the platoon leader in order to virtually extend the coverage of the platoon leader. A virtual leader selection algorithm is developed to designate a platoon member as a virtual leader effectively such that the system performance is maximized in terms of the stability for maintaining the desired inter-vehicle distance. More specifically, a new metric called the \emph{Virtual Leader Quality Index (VLQI)} is defined to quantify the effectiveness of a platoon member for serving as a virtual leader in terms of the capacity of extending the range and maintaining good network connectivity with the platoon leader. Additionally, we identify the problem that the standard mechanisms developed to handle vehicle join and leave maneuvers for platooning do not work efficiently for a long platoon, and we design novel approaches to managing the vehicle join and leave maneuvers specifically for a long platoon.

We perform extensive simulations using both the OMNeT++ based vehicular network simulation framework, Veins~\cite{sommer2019veins}, and a roadway traffic simulator, SUMO~\cite{behrisch2011sumo}. We demonstrate that \emph{L-Platooning} allows vehicles to form a long platoon effectively. Specifically, it is shown that \emph{L-Platooning} allows vehicles to maintain the inter-vehicle distance very precisely with the mean error of only 6cm. We also show that \emph{L-Platooning} effectively handles the vehicle join and leave maneuvers for a long platoon.

This paper is organized as follows. In Section~\ref{sec:background}, the background on platooning focusing on the control system and V2V communication technologies is presented. Based on the background, we describe the problem of the current platooning technology in Section~\ref{sec:problem}. We then present the details of the proposed protocol in Section~\ref{sec:proposed}. We perform simulations to evaluate the performance of \emph{L-Platooning} in Section~\ref{sec:evaluation}. Finally, we provide discussion on related issues and future research directions in Section~{\ref{sec:discussion}} and conclude in Section~{\ref{sec:conclusion}}.

\section{Background}
\label{sec:background}

A longitudinal drive control and V2V communication systems underpin the vehicle platooning technology. In this section, we present a brief review on these key components for platooning.

\subsection{Drive Control System for Platooning}
\label{sec:background_control}

\begin{figure}
\centering
\includegraphics[width=.99\columnwidth]{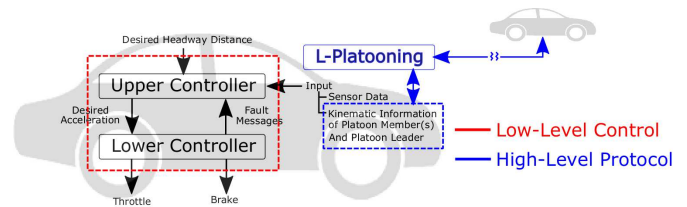}
\caption {An architecture of a platooning system integrated with the proposed approach.}
\label{fig:control_architecture}
\end{figure}

A longitudinal control system is comprised of an upper controller and a lower controller as shown in Fig.~\ref{fig:control_architecture}. The upper controller is a cruise controller that computes the acceleration of a vehicle to maintain desired inter-vehicle distance based on a set of inputs including on-board sensor data and kinematic data of platoon members. In particular, the kinematic information of other platoon members is received via V2V communication. The lower controller is used to control the actuation of a vehicle, \emph{i.e.,} the throttle and brake to achieve the desired acceleration.

The adaptive cruise control (ACC) has long been a traditional drive control system for platooning~{\cite{vahidi2003research}}. It automatically adjusts the acceleration in order to achieve the desired headway distance to the front vehicle based on local sensor data such as RADAR and LiDAR data, but it does not exploit the kinematic information of other platoon members that can be received wirelessly. Once the desired acceleration is determined, the lower controller adjusts throttle and brake to achieve the desired acceleration. Without utilizing the kinematic information of other platoon members, however, ACC fails to provide strong string stability~{\cite{rajamani2011vehicle}} where the string stability means that an error occurred in controlling a platoon member does not amplify as it propagates towards the end of the platoon.

The cooperative adaptive cruise control (CACC) algorithms are developed to address the problem of ACC, thereby providing stronger stability. Specifically, CACC utilizes not only the on-board sensor data but also the kinematic information of other platoon members received via V2V. Different kinds of CACC algorithms have been proposed in the literature~\cite{rajamani2000demonstration}\cite{ploeg2011design}\cite{di2014distributed}\cite{santini2015consensus}. In this paper, we employ one of the widely used CACC algorithms for platooning~{\cite{rajamani2011vehicle}}. Numerous vehicle controllers~{\cite{aramrattana2018evaluating}\cite{al2017feedforward}\cite{kianfar2012design}} are developed based on this algorithm or the performance was compared with this algorithm. Additionally, various platooning simulators such as~{\cite{segata2014plexe}\cite{segata2012simulation}\cite{jaworski2012microscopic}} are built upon this controller. Also, the PATH (Partners for Advanced Transportation Technology) and SARTRE (Safe Road Trains for the Environment) projects~{\cite{bergenhem2010challenges}\cite{rajamani2000demonstration}} are based on this type of CACC algorithm.

According to the CACC algorithm~\cite{rajamani2011vehicle}, the control law denoted by $\ddot{x}_{i\_des}$ for $i$-th platoon member to compute desired acceleration is calculated as follows.

\begin{equation}
\ddot{x}_{i\_des} = \alpha_1\ddot{x}_{i-1} + \alpha_2\ddot{x}_0 + \alpha_3\dot{\varepsilon}_i + \alpha_4(\dot{x}_i-\dot{x}_0)+\alpha_5\varepsilon_i
\end{equation}
\begin{equation}
\varepsilon_i = x_i - x_{i-1} + l_{i-1} + \mbox{gap}_{\mbox{des}}
\end{equation}
\begin{equation}
\dot{\varepsilon}_i = \dot{x}_i - \dot{x}_{i-1}
\end{equation}

\noindent Here $\ddot{x}_0$ and $\dot{x}_0$ are the acceleration and speed of the platoon leader; $\ddot{x}_{i-1}$ is the acceleration of the preceding vehicle; $l_{i-1}$ is the length of the preceding vehicle; $x_i$ and $x_{i-1}$ are the positions of the current vehicle and the preceding vehicle; thus $\varepsilon_i$ is the distance error with respect to $\mbox{gap}_{\mbox{des}}$ which is the desired inter-vehicle distance. The parameters $\alpha_i$ are defined as follows.

\begin{equation}
\alpha_1 = 1 - C_1; \mbox{ } \alpha_2 = C_1; \mbox{ } \alpha_5 = -\omega_n^2
\end{equation}
\begin{equation}
\alpha_3 = - (2\xi - C_1 ( \xi + \sqrt{\xi^2 - 1}))\omega_n
\end{equation}
\begin{equation}
\alpha_4 = - C_1 (\xi + \sqrt{\xi^2 - 1})\omega_n
\end{equation}

\noindent Here the default values for weighting factor $C_1$ between the acceleration of the preceding vehicle and that of the platoon leader, the damping ratio $\xi$, and the controller bandwidth $\omega_n$, which are used to determine the parameters $\alpha_i$, are summarized in Table~{\ref{table:setup}} according to~\cite{segata2014plexe}.

We observe that the upper controller fails to operate properly when the platoon leader is out of the range of DSRC since it does not receive the acceleration and speed of the platoon leader (\emph{i.e.,} $\ddot{x}_0$ and $\dot{x}_0$). The technical contribution of this paper thus lies in development of a platooning protocol and integration of the protocol with a vehicle control system to allow the controller to keep the constant inter-vehicle distance reliably regardless of the size of a platoon.

\subsection{Vehicle to Vehicle (V2V) Communication}

The V2V communication system is another important component for platooning. There are two types of V2V technologies for platooning: IEEE 802.11p (DSRC)~\cite{vukadinovic20183gpp} and 3GPP/LTE-based C-V2X~\cite{3gpptech}. While DSRC is based on the IEEE 802.11 standard, it has unique characteristics designed for fast moving vehicles. Specifically, the medium access control (MAC) layer of DSRC is a variation of the IEEE 802.11 standard to cope more effectively with dynamic vehicular environments~\cite{vukadinovic20183gpp}; DSRC eliminates the need of establishing a basic service set (BSS), thereby it executes in the Outside the Context of a BSS (OCB) mode with the carrier-sense multiple access with collision avoidance (CSMA/CA). The physical layer of DSRC is similar to IEEE 802.11a, while DSRC is amended based on orthogonal frequency-division multiplexing (OFDM) to facilitate communication among vehicles that move fast. More specifically, it introduces 10MHz channels rather than 20MHz channels in order to reduce the root-mean-squared delay spread in dynamic vehicular environments at the cost of cutting the maximum data rate down to 27Mb/s from 54Mb/s.


Another line of V2V communication technologies is the 3GPP/LTE-based C-V2X~\cite{3gpptech}. The third generation partnership project (3GPP) introduced Release 14 in 2016 which includes the V2V communication service. It supports both infrastructure-based, \emph{i.e.,} using eNodeB for resource allocation (Mode-3), and infrastructureless, also called as the sidelink/PC5 like DSRC, \emph{i.e.,} using a direct communication link between vehicles (Mode-4). For more details on C-V2X, readers are referred to the excellent review paper~\cite{naik2019ieee}.



It is widely expected that the two communication technologies will co-exist for some time. It is not only because of the automotive industry that is split in two, \emph{e.g.,}, Toyota, General Motors, and Volkswagen are the manufacturers of DSRC-equipped cars, while Ford, PSA Group, and Daimler have committed to the C-V2X technology. At the point of writing this paper, debates are going on in U.S., Europe, and China on determining the standard for V2V communication, and research on developing solutions that support both DSRC and C-V2X is very active. In the mean time, most practical platooning tests have been dominated by DSRC/WAVE~\cite{jiang2008ieee}, and its European counterpart ITS-G5~\cite{etsi2011intelligent}\cite{etsi2011intelligent2}. In this paper, simulations are conducted based on DSRC. However, it can be easily extended to the device-to-device communication mode for C-V2X for platooning.

\section{Problem Statement}
\label{sec:problem}


A main motivation of this work is that existing platooning solutions are designed without taking into account the limited range of V2V communication, thereby failing to manage and operate a possibly long platoon. Especially considering that the off-hours delivery is widely investigated and adopted by many cities~{\cite{wang2019optimal}\cite{sathaye2010unintended}}, a chance of utilizing a long platoon is expected to be high as the platooning technology is being adopted by freight companies, which creates the needs for a novel platooning protocol that can support seamless and effective formation of a long platoon despite the limited range.

\begin{figure}
\centering
\includegraphics[width=.9\columnwidth]{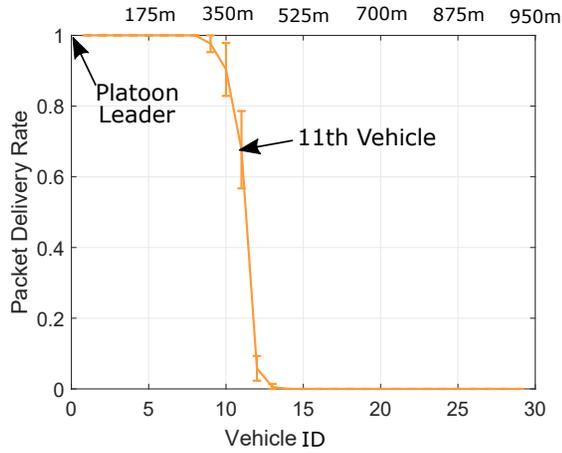}
\caption {PDR of platoon members. PDR starts to drop significantly from the 11th vehicle which is about 400m away from the leader.}
\label{fig:effect_of_dsrc_range_on_pdr}
\end{figure}

Recent measurement studies on DSRC demonstrate that the inter-vehicle distance and the signal propagation environment are key factors that determine the performance of DSRC~\cite{bai2010toward}\cite{lv2016empirical}. Especially, the packet delivery rate (PDR) for DSRC decreases significantly as the inter-vehicle distance increases. For example, in an urban environment, it was observed that PDR decreased from 93.6\% to 39.1\% as the inter-vehicle distance increased from 0m to 450m~\cite{bai2010toward}. We show that such a limited range of DRSC may interfere with an existing platooning solution in managing a long platoon in terms of maintaining a constant inter-vehicle distance to the front vehicle. Specifically, to understand better the effect of the limited range of DSRC on forming a long platoon, we perform simulation using Veins~\cite{sommer2019veins} and SUMO~\cite{behrisch2011sumo}. Veins is a framework for vehicular network simulation which is based on OMNeT++~\cite{varga2008overview}. SUMO~\cite{behrisch2011sumo} is a road traffic simulator. In particular, we adopt Plexe~\cite{segata2014plexe} a platooning extension for Veins. In this simulation study, a long platoon consisting of 30 vehicles (trailer trucks with body length of 13m) with inter-vehicle distance of 20m was created, and the platoon leader changed its speed continuously in a sinusoidal fashion to evaluate the string stability. The platoon members adjusted the inter-vehicle distance based on the CACC algorithm~{\cite{rajamani2011vehicle}}. Details on the simulation settings for V2V communication, vehicle mobility, and driving control system are described in Section~\ref{sec:simulation_setup}.

We measured PDR of each platoon member of the platoon. Results are shown in Fig.~\ref{fig:effect_of_dsrc_range_on_pdr}. In this simulation, vehicles within the range of 0$\sim$350m from the platoon leader (\emph{i.e.,} the first 8 platoon members) received a beacon message reliably from the leader with nearly 100\% PDR. On the other hand, PDR decreased significantly starting from the 11th vehicle of the platoon which is about 400m away from the leader which coincides with a previous measurement study~\cite{bai2010toward}. The PDR of the 12nd and 13rd vehicles were only about 5.8\% and 0.5\%, respectively, and all other following vehicles that are farther away from the leader than these vehicles had nearly 0\% PDR.

\begin{figure}
\centering
\includegraphics[width=.9\columnwidth]{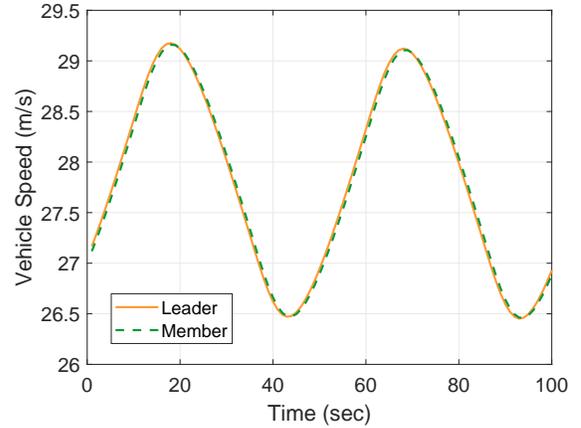}
\caption {Speed of a platoon member with high PDR and the platoon leader. The platoon member receives a beacon reliably from the platoon leader, thus precisely adjusting the speed according to that of the leader that changes in a sinusoidal fashion.}
\label{fig:leader_speed}
\end{figure}

\begin{figure}
\centering
\includegraphics[width=.9\columnwidth]{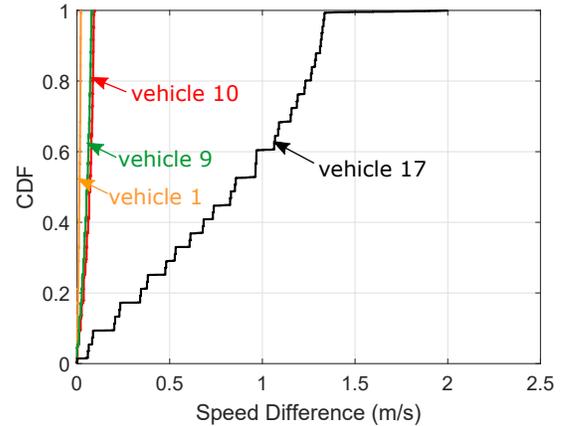}
\caption {Speed difference compared with that of the platoon leader. Vehicle 17 that does not receive a beacon message from the platoon leader reliably fails to adjust its speed according to that of the platoon leader.}
\label{fig:effect_of_dsrc_range_on_speed}
\end{figure}

\begin{figure*}
\centering
\includegraphics[width=.9\textwidth]{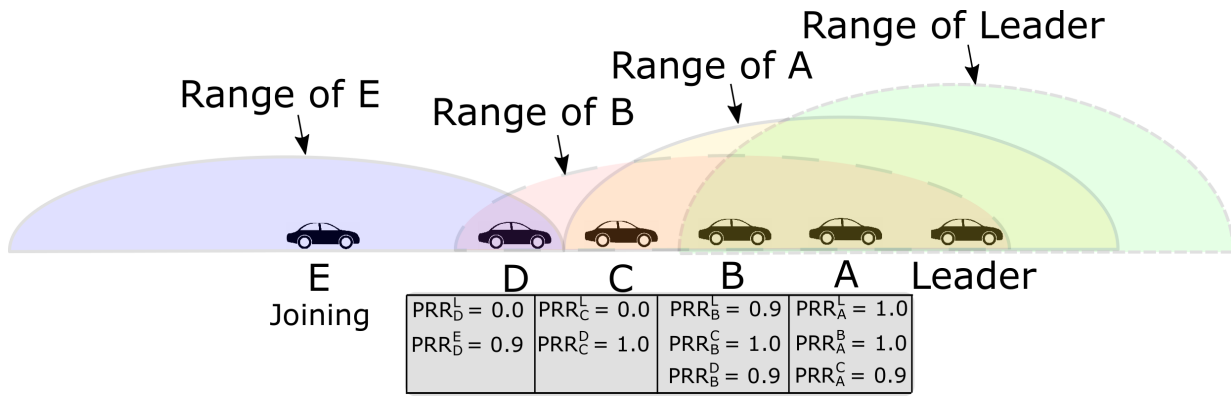}
\caption {Illustration of VLQI calculation. Vehicle $B$ has more followers that have very poor connection with the platoon leader than vehicle $A$. Thus, it is selected as the virtual leader, although vehicle $A$ has slightly better connection with the platoon leader.}
\label{fig:how_it_works}
\end{figure*}

The platoon members that do not receive a beacon message from the platoon leader reliably fail to adjust their acceleration appropriately and do not maintain the constant inter-vehicle distance. Fig.~\ref{fig:leader_speed} shows how a platoon member with a high PDR adjusts its speed in accordance with the speed change of the platoon leader. It is shown that due to the high PDR, the vehicle accurately adjusts its speed in a sinusoidal fashion according to the speed change of the platoon leader. On the other hand, platoon members with a low PDR fail to adapt their speed correctly as shown in Fig.~\ref{fig:effect_of_dsrc_range_on_speed}. Specifically, Fig.~\ref{fig:effect_of_dsrc_range_on_speed} displays the speed difference compared to that of the platoon leader for four platoon members, \emph{i.e.,} 1st, 9th, 10th, and 17th vehicles. We observe that the 1st vehicle has nearly 0 speed difference, and as the distance from the platoon leader increases, the speed difference increases slightly; and the 17th vehicle with 0 PDR completely fails to adjust its speed.

\begin{table}
\begin{center}
\begin{tabular}{ |l|l|l| }
\hline
\textbf{Paper} & \textbf{Join} & \textbf{Leave} \\ \hline
\cline{1-3}\hline
Amoozadeh~\cite{amoozadeh2015platoon} & Msg to Platoon Leader & Msg to Platoon Leader \\ \hline
Segata~\cite{segata2014supporting} & Msg to Platoon Leader & Not Implemented \\ \hline
Kazerooni~\cite{kazerooni2015interaction} & Msg to Platoon Leader & Not Implemented \\ \hline
Santini~\cite{santini2018platooning} & Msg to Platoon Leader & Msg to Platoon Leader \\ \hline
Ploeg~\cite{ploeg2017cooperative} & Msg to Platoon Leader & Not Implemented \\ \hline
\end{tabular}
\caption{A summary of protocols for join/leave maneuvers for platooning.}
\end{center}
\end{table}

In this paper, we aim to address this problem and develop a novel platooning protocol that allows vehicles to form a long platoon seamlessly and effectively. Specifically, we propose a platooning protocol based on a novel concept called the virtual leader that assists with the platoon leader in managing a long platoon so that the platoon members can maintain the constant inter-vehicle distance with string stability regardless of the size of a platoon (Section~{\ref{subsec:overview}}) In this protocol, we address the problem of how to select such a virtual leader effectively (Section~{\ref{subsec:select}}) and share the selection of virtual leaders with the platoon members efficiently (Section~{\ref{subsec:consensus}}) Additionally, we develop novel schemes for supporting the vehicle join and leave maneuvers for a long platoon (Section~{\ref{subsec:handle_join}}) Numerous research has been conducted to create protocols to support the join and leave maneuvers for platooning~{\cite{amoozadeh2015platoon}\cite{segata2014supporting}\cite{kazerooni2015interaction}\cite{santini2018platooning}\cite{santini2018platooning}\cite{ploeg2017cooperative}}. The key characteristics of such protocols are summarized in Table I. As shown in this table, these protocols are mostly based on exchanging a ``request'' message with the platoon leader which may not work if the platoon leader is out of the range of DSRC or D2D for C-V2X. In particular, in this paper, we aim to develop a higher-level platooning protocol to support seamless formation of a long platoon regardless of the platoon size which can be integrated with any lower-level controllers. As such, we assume that the platoon is equipped with an adequate lower-level controller (\emph{i.e.,} the default lower level controller of plexe~{\cite{segata2014plexe}}).

\section{L-Platooning}
\label{sec:proposed}

We present an overview of the \emph{L-Platooning} followed by details of each component of the proposed protocol.

\subsection{Overview}
\label{subsec:overview}

\emph{L-Platooning} is developed to support seamless formation of a platoon regardless of the size of a platoon while sustaining the string stability and to handle effectively the vehicle join and leave maneuvers for a long platoon. A simple approach to allow platoon members that are out of the communication range of the platoon leader to receive a DSRC message from the platoon leader is to make platoon members piggyback the message in its beacon message and broadcast it in the next beacon interval. This simple approach is not a viable option for managing a long platoon. More specifically, in DSRC, time is divided into 100ms sync periods which consist of a 50ms control channel (CCH) interval and a 50ms service channel (SCH) interval~\cite{kenney2011dedicated}. Thus, it will take up to 100ms for a platoon member to relay the piggybacked message. As such, in order to relay the DSRC message from the platoon leader to a platoon member that is $n$ hops away from the platoon leader, the maximum latency would be $100 * n$ ms, which causes a significantly delay in maintaining the inter-vehicle distance accurately and consistently.

Our approach is based on a novel idea of designating platoon member(s) to serve as ``virtual'' platoon leader(s), so that the virtual platoon leader can directly communicate with its platoon members that are out of the communication range of the original platoon leader, without relying on multi-hop communication. Specifically, we introduce a new concept called the \emph{Virtual Leader} which is a platoon member that is selected to act as an intermediate platoon leader. More precisely, the following vehicles of a virtual leader consider the virtual leader as the original platoon leader and use the speed and acceleration of the virtual leader in adjusting the inter-vehicle distance based on the drive control system.

Basically, the key mechanism of \emph{L-Platooning} is to divide a long platoon into multiple manageable small platoons by electing virtual leaders that serve each of those small platoons. \emph{L-Platooning} is comprised of a number of protocol components: virtual leader section algorithm (Section~\ref{subsec:select}), virtual leader assignment algorithm (Section~\ref{subsec:consensus}), and an algorithm to handle the vehicle join and leave maneuvers for a long platoon (Section~\ref{subsec:handle_join}). Details of these protocol components are described in the following sections.

\subsection{Selecting Virtual Leaders}
\label{subsec:select}

In this section, we explain how \emph{L-Platooning} selects a virtual leader. The basic idea of the proposed virtual leader selection algorithm is to select a platoon member as a virtual leader that has good network connectivity with both the platoon leader and its following vehicles. Specifically, each platoon member monitors and quantifies the quality of the connectivity, and sends that information to the platoon leader. Consequently, the platoon leader, upon receiving the beacon messages from its platoon members, selects a virtual leader based on the connectivity information contained in the messages.

Each vehicle uses a real-time link quality estimator to quantify the quality of the connectivity. Numerous real-time link quality estimators have been developed~\cite{baccour2012radio}\cite{vlavianos2008assessing}. Especially for \emph{L-Platooning}, we adopt the exponentially weighted average packet reception ratio (EWMPRR)~\cite{woo2003evaluation} to quantify the quality of the connectivity. EWMPR is a very simple and memory efficient mechanism for quantifying the link quality, requiring only 2 multiplications and 1 addition. As such, it is highly applicable to our application where vehicle platooning requires very frequent message transmissions and rapid processing. However, note that EWMPR can be easily replaced with a different link quality estimator depending on the application specific needs.

The link quality with respect to platoon member $j$ measured by vehicle $i$ is denoted by $PRR_i^{j}$. In particular, the link quality with respect to the platoon leader measured by vehicle $i$ is denoted by $PRR_i^L$. Now the degree of connectivity between vehicle $i$ and its following vehicle(s) is defined as follows.

\begin{equation}
\label{eq:7}
CON_i^{Follow}= \sum_{j \in \mbox{following vehicles}} (PRR_i^j - PRR_j^L).
\end{equation}

Eq.~{\ref{eq:7}} indicates that two factors are considered in determining a virtual leader: (1) level of connectivity with its following vehicles ($PRR_i^j$) and (2) level of connectivity between the following vehicles and their leader ($PRR_j^L$). According to this equation, an ideal virtual leader would be the one that covers as many following vehicles as possible that have poor connectivity with their leader (\emph{i.e.,} the following vehicles that require a new leader due to poor connectivity).


Now we are ready to define the \emph{Virtual Leader Quality Indicator (VLQI)} of vehicle $i$ which is denoted by $VLQI_i$. Since an ideal virtual leader maintains good connectivity with both its following vehicles (\emph{i.e.,} represented by the metric $CON_i^{Follow}$) and its platoon leader which is denoted by $CON_i^{Leader}$, the VLQI of vehicle $i$ is defined as follows.

\begin{equation}
VLQI_i = \gamma \times CON_i^{Leader} + (1 - \gamma) \times CON_i^{Follow}.
\end{equation}

\noindent Here $CON_i^{Leader}$ is simply $PRR_i^L$, \emph{i.e.,} it is the EWMPR with respect to the platoon leader for vehicle $i$, and $\gamma$ is a system parameter that is used to determine the weight between the connectivity with the following cars and the platoon leader. In our experiments, we used $\gamma = 0.5$.

Each vehicle $i$ calculates the virtual leader quality indicator $VLQI_i$ and then piggybacks the result in its beacon message. As a result, the platoon leader receives the VLQI values from its platoon members and selects a platoon member with the largest VLQI value as a virtual leader. Here we present a simple working example to help readers better understand our virtual leader selection algorithm. Consider Fig.~\ref{fig:how_it_works} for an example. The platoon leader receives beacon messages from vehicles $A$ and $B$. Thus, one of these two vehicles is selected as the virtual leader. The VLQI value for vehicle $A$ is $0.5 \cdot 1.0 + 0.5 \cdot \{(1.0-0.9) + (0.9-0.0)\} = 1.0$, and the VLQI value for vehicle $B$ is $0.5 \cdot 0.9 + 0.5 \cdot \{(1.0-0.0) + (0.9-0.0)\} = 1.4$. In this example, although vehicle $A$ has better connection with the platoon leader than vehicle $B$, since vehicle $B$ covers more followers that have very poor connection with the platoon leader ($PRR_C^L=0$ and $PRR_D^L=0$), vehicle $B$ is selected as the virtual leader. Essentially, the VLQI model indicates in this example that the advantage of covering more followers is greater than having slightly better connection with the platoon leader ($PRR_B^L=0.9$ and $PRR_A^L=1.0$).

\begin{figure}
\centering
\includegraphics[width=.9\columnwidth]{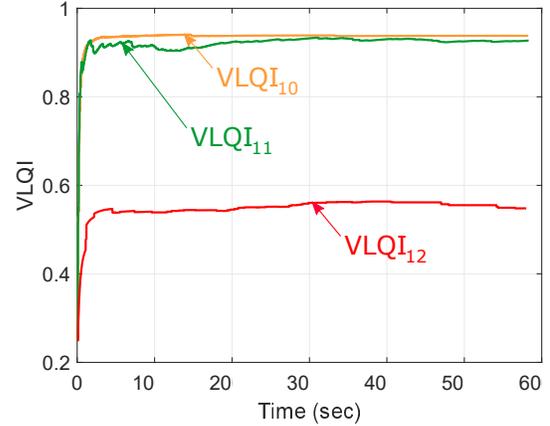}
\caption {VLQI values for vehicles 10, 11, and 12 measured over time. The VLQI value of each vehicle quickly converges, being ready to be used by the platoon leader for selecting vehicle 10 as a virtual leader.}
\label{fig:demo_vlqi}
\end{figure}

We also show through simulations how VLQI values change and consequently converge reliably throughout the operation of platooining consisting of 30 trucks based on the same simulation settings as described in Section~\ref{sec:problem}. Fig~\ref{fig:demo_vlqi} shows the results. As shown, the VLQI values of all vehicles quickly converged. In this simulation, vehicle 10 is selected as the virtual leader because it has the highest VLQI value. Although vehicle 11 had more followers with poor connectivity with the platoon leader than vehicle 10, in this case, $PRR_{11}^L$ is too small compared with vehicle 10 (See Fig.~\ref{fig:effect_of_dsrc_range_on_pdr}). Similarly, vehicle 12 is not selected as the virtual leader because it has very poor connectivity with the platoon leader, thereby having a very small VLQI value.

\subsection{Updating Virtual Leaders for Platoon Members}
\label{subsec:consensus}

\begin{figure}
\centering
\includegraphics[width=.99\columnwidth]{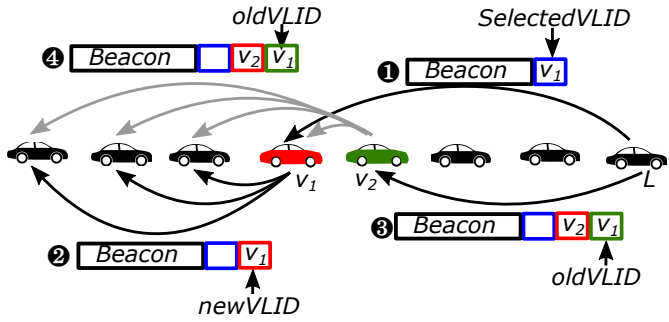}
\caption {A diagram illustrating the process of updating the virtual leader.}
\label{fig:vl_update_diagram}
\end{figure}

Once a virtual leader is selected, the following vehicles of the virtual leader should update their platoon leader to the newly selected virtual leader. This process of updating the virtual leader is illustrated in Fig.~{\ref{fig:vl_update_diagram}}. First, the original platoon leader notifies the selected vehicle that it is selected as a virtual leader. This can be simply implemented by putting the ID of the newly selected virtual leader in a new 4 byte field \emph{selectedVLID} in a DSRC beacon message
({\circled{1}}), \emph{i.e.,} when a platoon member receives a beacon message from the platoon leader, the platoon member sets itself as a virtual leader if its ID is the same as the one specified in the \emph{selectedVLID} field. The virtual leader then notifies its following vehicles to update their platoon leader to the virtual leader. This is done by adding another 4 byte field \emph{newVLID} in the DSRC beacon message ({\circled{2}}). Upon receiving a beacon message from the new virtual leader, the following vehicles will set their platoon leader to the newly selected virtual leader. From this time on, the following vehicles start to calculate their VLQI values based on the new virtual leader. In addition, the new virtual leader can select another virtual leader subsequently if necessary based on the VLQI values received from its following vehicles. This way \emph{L-Platooning} effectively manages a long platoon by selecting virtual leaders subsequently and letting the virtual leaders to handle their own following vehicles.

A technical challenge arises when a virtual leader should be re-elected if the previous virtual leader leaves from the platoon and/or the link quality changes, \emph{i.e.,} a platoon member with better VLQI values appears. The strength of \emph{L-Platooning} is the capability of adaptively updating the virtual leader by keeping monitoring the VLQI values of their following vehicles and selecting a new virtual leader automatically. When a virtual leader is changed, the original leader puts the ID of the previous virtual leader in the new 4 byte field \emph{oldVLID} and the ID of the new virtual leader in the field \emph{newVLID}, and broadcast the beacon message ({\circled{3}}). Upon receiving this beacon message, the old virtual leader becomes a regular platoon member stopping the service as the virtual leader, and the vehicle with the new virtual leader ID becomes the new leader. Both the new and old virtual leaders keep broadcasting the beacon message with the ID of the previous virtual leader in the \emph{oldVLID} field so that any following vehicle can update their leader to the new virtual leader ({\circled{4}}). Additionally, to prevent frequent changes of the virtual leader especially when VLQI values change significantly in the beginning of platoon formation due to the highly fluctuating nature of EWMPR, we ensure that the platoon leader selects a virtual leader if it has the highest VLQI value in the $\beta$ consecutively received beacons.

Once a vehicle is updated with a new virtual leader, the vehicle starts to use directly the speed and acceleration data of its virtual leader, making the driving control process (\emph{i.e.,} calculating acceleration to achieve the desired inter-vehicle distance) extremely fast. For example, in Fig.~\ref{fig:how_it_works}, vehicle $B$ is selected as the virtual leader, and its following vehicles $C$ and $D$ use the speed and acceleration of vehicle $B$ in calculating their acceleration to maintain constantly small inter-vehicle distance.

\begin{figure}
\centering
\includegraphics[width=.9\columnwidth]{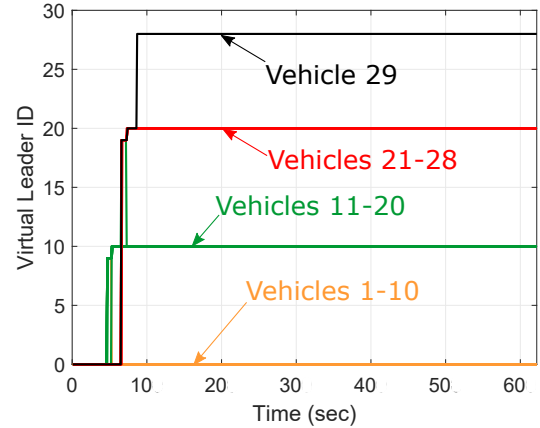}
\caption {Selection of virtual leader. All vehicles are assigned with either a virtual leader or a original platoon leader ($\beta=5$).}
\label{fig:leaderSelection}
\end{figure}

We verify that the virtual leader selection algorithm works correctly and all vehicles are updated with new virtual leaders in a timely manner through simulations. The same simulation settings ($\beta = 5$) are used as described in Section~\ref{sec:problem}. The results are shown in Fig.~\ref{fig:leaderSelection}. It is shown that in this simulation setting, 3 virtual leaders are selected to serve a platoon consisting of 30 trailer trucks with body length of 13m and the desired inter-vehicle distance of 20m. It is interesting to note that there is a virtual leader near the tail of the platoon, \emph{i.e.,} 28th vehicle serving 29th vehicle and being ready to take any platoon join request from vehicles that wish to join the platoon. We will discuss this great property of \emph{L-Platooning} that allows vehicles seamlessly join and leave from a `long' platoon in the following section. The results also demonstrate another strong aspect of \emph{L-Platooning}, \emph{i.e.,} selection and update of virtual leaders are completed very rapidly, in this example, within 10 seconds.

While \emph{L-Platooning} is very effective and fast protocol for managing a long platoon due to the simple distributed mechanisms for selecting a virtual leader and updating the virtual leader of platoon members, it has some overhead in terms of the increased beacon message size. More specifically, the additional message overhead is 28 bytes to add the new fields, $VLQI_i$, $PRR_i^L$, \emph{selectedVLID, newVLIDI, and oldVLID}. Empirical studies, however, show that a packet size increase of 100bytes has only slight impact on performance~\cite{yin2004performance}. We expect that an increase of only 28 bytes will have a very marginal effect. We analyze the impact of the increased packet size in more detail in Section~\ref{sec:packet_size}.

\subsection{Handling Vehicle Join/Leave Maneuvers}
\label{subsec:handle_join}

In this section, we describe how \emph{L-Platooning} handles the vehicle join and leave maneuvers. A standard approach for managing the vehicle join maneuver is to allow a joining vehicle to send a request message to the platoon leader; and then the platoon leader allows the vehicle to join by sending a reply message~\cite{segata2014plexe}. Similarly, a vehicle can leave from a platoon by sending a request message to the platoon leader and receiving a reply from it~\cite{segata2014plexe}. However, we note that this standard approach for handling the vehicle join and leave maneuvers do not work for a long platoon because the platoon leader may fail to receive the request message when the joining/leaving vehicle is out of the DSRC range of the platoon leader.

\emph{L-Platooning} addresses this problem for a long platoon by allowing a joining/leaving vehicle to communicate directly with their virtual leader. An interesting aspect of \emph{L-Platooning} is that there is always a virtual leader near the end of a platoon that can receive a request from a joining vehicle. This property can be simply proved based on the proof by contradiction. Assume in contradiction that a joining vehicle does not have connection with a virtual leader, \emph{i.e.,} the joining vehicle is out of range of the virtual leader. This is contradiction because the virtual leader must have selected one of the following vehicles within its range as another virtual leader. Consider Fig.~\ref{fig:how_it_works} for an example. Here vehicle $E$ is a joining vehicle. Although vehicle $E$ is close to vehicle $D$, it does not have connection with vehicle $B$ which is the virtual leader. Since vehicle $B$ would have selected vehicle $D$ or $C$ as the virtual leader according to the protocol, the joining vehicle $E$ should be able to join the platoon.

\emph{L-Platooning} handles elegantly the vehicle leave maneuver as well. Specifically, there are two possible vehicle leaving scenarios: (1) a leaving vehicle is not a virtual leader, and (2) a leaving vehicle is a virtual leader. The former case can be handled simply. More specifically, when a vehicle leaves, the immediately following vehicle closes the gap based on the control system, and then its virtual leader recalculates the VLQI values for its following vehicles and re-select a virtual leader if necessary. The second scenario is a little bit trickier because when a virtual leader leaves, its following vehicles no longer receive a beacon message from the leader which has left already. To address this challenge, \emph{L-Platooning} ensures that if a leaving vehicle is a virtual leader, it first sends a leave request to its immediate follower and designate the follower as the new virtual leader. If it does not have a follower, it can just simply leave. After the vehicle leaves, the new virtual leader closes the gap starting to serve its following vehicles. In Sections~\ref{sec:vehicle_join_performance} and~\ref{sec:vehicle_leaving_performance}, we demonstrate that these mechanisms of \emph{L-Platooning} for handling the vehicle join and leave maneuvers work effectively.

\section{Evaluation}
\label{sec:evaluation}

\subsection{Simulation setup}
\label{sec:simulation_setup}

We consider a platoon consisting of 30 trailer trucks with body length of 13m led by a platoon leader which continuously changes its speed in a sinusoidal fashion. Note that the benefits of \emph{L-Platooning} are clear even for a smaller platoon consisting of, \emph{e.g.,} 10 to 15 trucks, but to demonstrate that \emph{L-Platooning} effectively reorganizes a long platoon into multiple ``sub-platoons'' with multiple virtual leaders, we created a platoon with 30 trucks in this simulation. Specifically, \emph{L-Platooning} was implemented in a vehicular simulation framework Veins/Plexe~\cite{sommer2019veins}\cite{segata2014plexe} incorporated with a traffic simulator SUMO~\cite{behrisch2011sumo}. The desired inter-vehicle distance is set to 20m with $\gamma=0.5$ and $\beta=5$. All other simulation parameters for V2V communication, car mobility, and driving controller are summarized in Table~
\ref{table:setup}.

\begin{table}
    \caption{Simulation Parameters for V2V Communication, Mobility, and Controller}
    \label{table:setup}
    \center
\begin{tabular}{c|l|l|}
        \cline{2-3}
        & Parameter & Value \\ \cline{1-3}
\multicolumn{1}{ |c|  }{} & Path loss model & Free space \\
\multicolumn{1}{ |c|  }{} & PHY model & IEEE 802.11p \\
\multicolumn{1}{ |c|  }{} & MAC model & IEEE 1609.4 \\
\multicolumn{1}{ |c|  }{} & Frequency & 5.89GHz \\
\multicolumn{1}{ |c|  }{} & Bitrate & 6 Mbit/s (QPSK R = $\frac{1}{2}$) \\
\multicolumn{1}{ |c|  }{} & Access category & AC\_VI \\
\multicolumn{1}{ |c|  }{} & Thermal noise & -85dBm \\
\multicolumn{1}{ |c|  }{} & Packet size & 228Byte \\
\multicolumn{1}{ |c|  }{\rot{\rlap{~Communication}}} & TX power & 20dBm \\ \cline{1-3}
\multicolumn{1}{ |c|  }{} & Leader's average speed & 100km/h  \\
\multicolumn{1}{ |c|  }{} & Oscillation frequency & 0.2Hz  \\
\multicolumn{1}{ |c|  }{} & Oscillation amplitude & $\simeq$  95 km/h to 105 km/h \\
\multicolumn{1}{ |c|  }{} & Platoon size & 30 cars  \\
\multicolumn{1}{ |c|  }{\rot{\rlap{~Mobility}}} & Car length & 13m (Truck) \\ \cline{1-3}
\multicolumn{1}{ |c|  }{} & Engine lag $\tau$ & 0.5s  \\
\multicolumn{1}{ |c|  }{} & Weight factor $C_1$ & 0.5  \\
\multicolumn{1}{ |c|  }{} & Controller bandwidth $\omega_n$ & 0.2Hz  \\
\multicolumn{1}{ |c|  }{} & Damping factor $\xi$ & 1  \\
\multicolumn{1}{ |c|  }{} & Desired gap gap$_{des}$ & 20m  \\
\multicolumn{1}{ |c|  }{} & Headway time $T$ & 0.3s and 1.2s  \\
\multicolumn{1}{ |c|  }{} & ACC paramter $\lambda$ & 0.1  \\
\multicolumn{1}{ |c|  }{} & Distance gain $k_d$ & 0.7  \\
\multicolumn{1}{ |c|  }{} & Speed gain $k_s$ & 1.0  \\
\multicolumn{1}{ |c|  }{\rot{\rlap{~Controller}}} & Desired speed $\dot{x}_{des}$ (followers) & 130km/h  \\ \cline{1-3}
\end{tabular}
\end{table}

In this simulation, we focus on measuring the inter-vehicle distance of each platoon member in evaluating the performance of \emph{L-Platooning} because the key objective of employing the platooning technology is to maintain an inter-vehicle distance constantly. More specifically, we examine if each vehicle accurately maintains the desired inter-vehicle distance (20m) when the speed of the platoon leader continuously changes (Section~\ref{sec:intervehicle_performance}). We then measure the amount of time that \emph{L-Platooning} takes to select virtual leaders and update the virtual leaders of the corresponding platoon members (Section~\ref{sec:vlselection_performance}). We also analyze the effect of the increased packet size considering that \emph{L-Platooining} requires a few additional fields in the DSRC beacon message (Section~\ref{sec:packet_size}). Finally, we study the effectiveness of \emph{L-Platooning} in terms of handling the vehicle join and leave maneuvers. (Sections~\ref{sec:vehicle_join_performance} and~\ref{sec:vehicle_leaving_performance}) focusing on the delay for completing the vehicle join and leave requests.

\subsection{Inter-Vehicle Distance}
\label{sec:intervehicle_performance}

\begin{figure*}
\centering
\includegraphics[width=.99\textwidth]{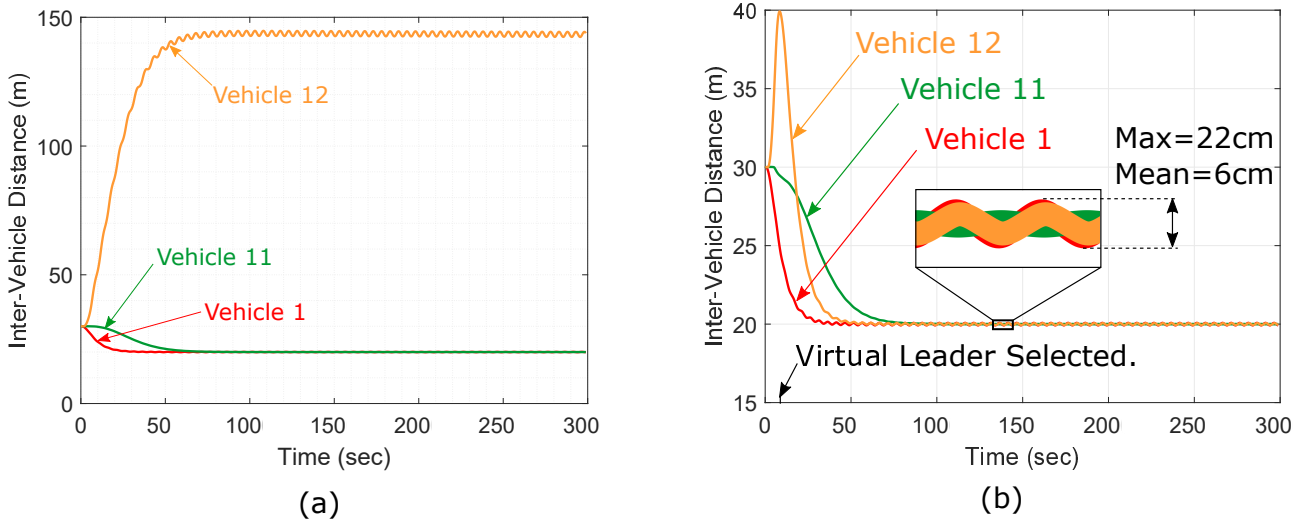}
\caption {(a) the inter-vehicle distance before applying \emph{L-Platooning}. Vehicle 12 fails to maintain the desired inter-vehicle distance because it does not receive a DSRC beacon from the platoon leader; (b) the inter-vehicle distance after applying \emph{L-Platooning}.}
\label{fig:dist_to_front_vehicle_combined}
\end{figure*}

In this section, we evaluate the performance of \emph{L-Platooning} in terms of the inter-vehicle distance maintained by each platoon member. Fig.~\ref{fig:dist_to_front_vehicle_combined}(a) illustrates the inter-vehicle distance of vehicles 1, 11, and 12 as a function of time before applying \emph{L-Platooning}. The graph shows that vehicles 1 and 11 accelerated quickly to reduce the inter-vehicle distance down to the desired value of 20m and accurately maintained the desired inter-vehicle distance, because these vehicles received the DSRC messages from both the preceding vehicle and the platoon leader reliably (Vehicle 0).

An interesting observation is that vehicle 1 achieved the desired inter-vehicle distance earlier than vehicle 11. The reason is because vehicle 1 is geographically closer to the platoon leader than vehicle 11. More specifically, we observed that platoon members achieve the inter-vehicle distance only after their front vehicles have attained the desired inter-vehicle distance because the driving control system depends on the speed and acceleration of the preceding vehicle. On the other hand, we observed that vehicle 12 failed to receive a beacon message from the platoon leader and as such it did not accelerate while keeping the same speed based on the default cruise control mode. As a result, the inter-vehicle distance for vehicle 12 increased until all vehicles finished adjusting their inter-vehicle distance to 20m. And then, the speeds of the front vehicles 1$\sim$11 were synchronized with that of the platoon leader. Consequently, the inter-vehicle distance of vehicle 12 fluctuated in a sinusoidal fashion as the speed of the front vehicles continuously changes according to that of the platoon leader.


We then applied \emph{L-Platooning} and measured the inter-vehicle distances. The results are displayed in Fig.~\ref{fig:dist_to_front_vehicle_combined}(b). All vehicles successfully adjusted their inter-vehicle distances to 20m. In particular, vehicle 12 was also able to adjust its distance to 20m because the virtual leader (vehicle 10) was selected and assigned to vehicle 12 at about 15sec. Note that the inter-vehicle distance of vehicle 12 increased in the beginning of the simulation because of the delay to select and assign the virtual leader. Overall, \emph{L-Platooning} allowed all vehicles to precisely keep the inter-vehicle distance of 20m, although the distance fluctuated slightly because the leader continuously changed its speed. Despite the continuous speed change of the platoon leader, the inter-vehicle distance was kept nearly constant with the mean and max error of only 6cm and 22cm, respectively. To get a better sense of the obtained error of the inter-vehicle distance, we compare the error with other works: Huang \emph{et al.}~{\cite{huang2016toward}} (up to 20cm), Ali \emph{et al.}~{\cite{ali2015urban}} (up to 25cm), Wei \emph{et al.}~{\cite{wei2017event}} (up to 15cm), Gao \emph{et al.}~{\cite{gao2016robust}} (up to 18cm), and Zou \emph{et al.}~{\cite{zou2019event}} (up to 15.6cm). Considering that the error of \emph{L-Platooning} is measured for a long platoon with the constraint of the limited range of DSRC, while other works are measured for a regular platoon, the obtained error can be deemed quite competitive.

\subsection{Delay for Platooning}
\label{sec:vlselection_performance}

An interesting question is to understand how long it will take for \emph{L-Platooning} to allow all platoon members to achieve the desired inter-vehicle distance and start to maintain it precisely. To answer this question, in this section, we measured the delay required to reach the ``synchronized'' state. Specifically, we repeated simulations 100 times with different random seeds and recorded the delay for each platoon member.

\begin{figure}
\centering
\includegraphics[width=.9\columnwidth]{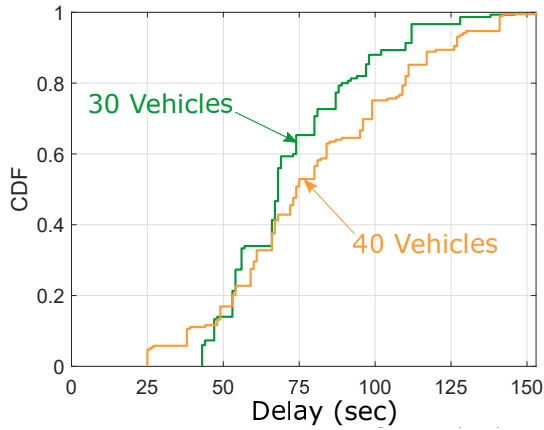}
\caption {Delay for completing the virtual leader selection algorithm and updating the virtual leaders of all platoon members. It is shown that the longer platoon takes about 10\% longer time to complete this process.}
\label{fig:time_for_vl_selection}
\end{figure}

Figure~\ref{fig:time_for_vl_selection} shows the results for two different platoon sizes of 30 and 40 trucks in the form of the cumulative distribution function (CDF) of the measured delay. It is straightforward from the graph that forming a longer platoon takes more time because longer platoons require more virtual leaders to be selected, and more time is needed to allow platoon members update their virtual leaders accordingly. More accurately, the average delay for the platoon with 30 vehicles was 7.2 seconds, while that for the platoon with 40 vehicles was 7.9 seconds. Despite the differences, it should be noted that \emph{L-Platooning} completes the virtual leader selection and the process of updating the virtual leader of platoon members very quickly for a very large platoon consisting of 30 to 40 trailer trucks.

\subsection{Effect of Packet Size}
\label{sec:packet_size}

\begin{figure}
\centering
\includegraphics[width=.9\columnwidth]{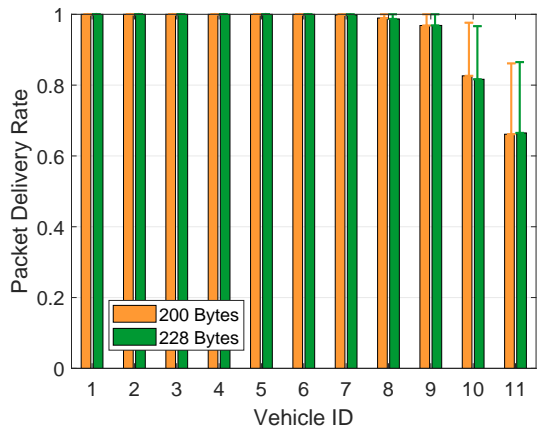}
\caption {Effect of increased packet size (28bytes). No statistically significant performance degradation is found due to the increased packet size.}
\label{fig:packet_size}
\end{figure}

\emph{L-Platooning} requires to add additional fields (\emph{i.e.,} additional 28 bytes) in a DSRC beacon message increasing the packet size. In this section, we investigate the impact of the increased packet size focusing on the packet delivery rate for each platoon member.

Fig.~\ref{fig:packet_size} shows the packet delivery rates for different packet sizes. The results indicate that there is no statistically significant difference in terms of the packet delivery rate between the two packet sizes. In fact, the literature on DSRC shows that even an increase of 100 bytes for a DSRC beacon message does not have significant impact on the performance~\cite{yin2004performance}.

\subsection{Vehicle Join Maneuver}
\label{sec:vehicle_join_performance}

To evaluate the performance of \emph{L-Platooning} in terms of how it handles the vehicle join maneuver, we create a scenario in which a vehicle joins after a long platoon has been formed. The speed of the joining vehicle is set to be higher than the speed of the platoon so that the joining vehicle can quickly catch up with the platoon and join. Once the vehicle catches up with the platoon, \emph{i.e.,} the vehicle is geographically close to the tail of the platoon, the vehicle sends a join request. This request message will be received by the closest virtual leader of the platoon. In this simulation, we varied the time when the request message is sent, \emph{i.e., } when the distance between the joining vehicle and the last vehicle of the platoon is 100m, 150m, 200m, and 250m. We then measured the amount of time \emph{L-Platooning} takes to complete the join maneuver.

\begin{figure*}
\centering
\includegraphics[width=.95\textwidth]{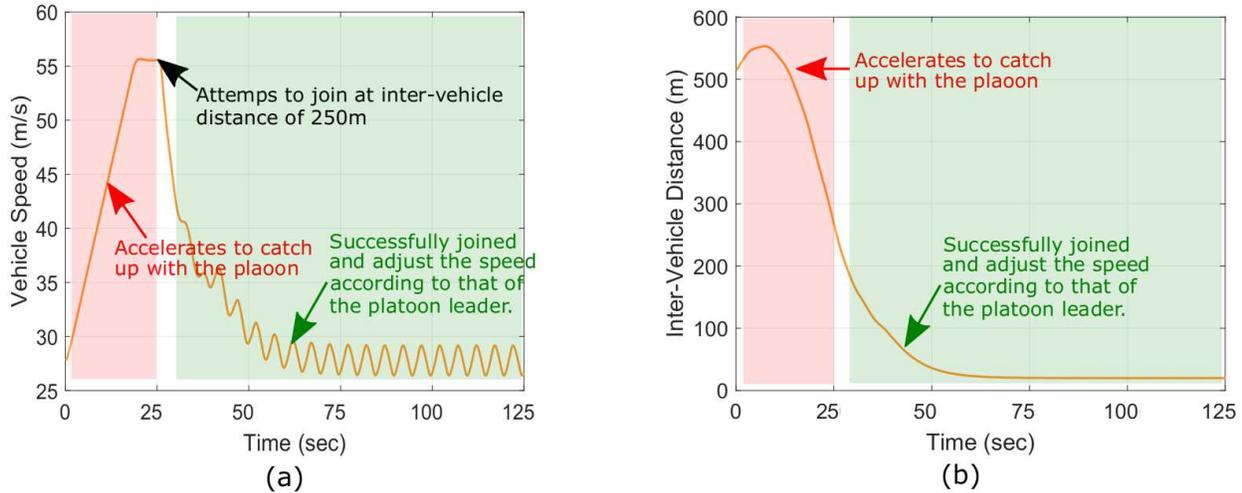}
\caption {(a) The speed of the joining vehicle. It reduces the speed once it catches up with the platoon and receives a reply from the closest virtual leader; (b) The inter-vehicle distance of the joining vehicle. The vehicle has successfully joined and maintained constantly the inter-vehicle distance of 20m.}
\label{fig:join-speed_combined}
\end{figure*}


Fig.~\ref{fig:join-speed_combined}(a) shows the speed of the joining vehicle. It is shown that the joining vehicle increases the speed to catch up with the platoon in the beginning of the simulation. Once the vehicle is close to the tail of the platoon, it sends a request message to the closest virtual leader and attempts to join the platoon. As the vehicle receives a reply message from the virtual leader, the joining vehicle significantly decreases the speed to adjust the inter-vehicle distance to 20m, completing the join maneuver. After that, the speed of the joining vehicle changes according to that of the platoon leader which fluctuates in a sinusoidal fashion.

Fig.~\ref{fig:join-speed_combined}(b) displays the inter-vehicle distance of the joining vehicle. It is shown that the inter-vehicle distance quickly drops as the joining vehicle increases its speed to catch up with the platoon. Once the vehicle is close enough to the front vehicle, \emph{i.e.}, at about 25sec, it attempts to join the platoon. Once the vehicle is joined, it starts to gradually reduce the inter-vehicle distance and then achieves the inter-vehicle distance of 20m.

\begin{figure}
\centering
\includegraphics[width=.9\columnwidth]{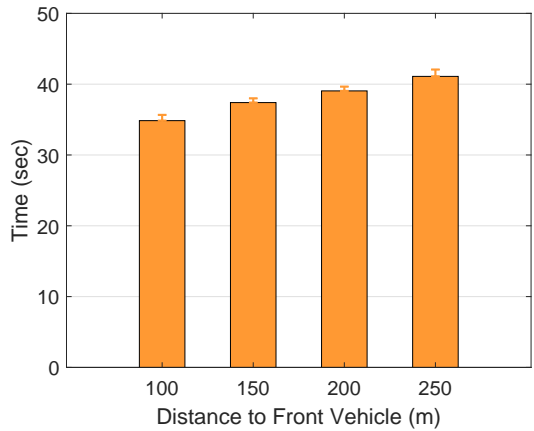}
\caption {The delay for \emph{L-Platooning} to complete the vehicle join maneuver for different distances between the joining vehicle and the front vehicle. It is shown that \emph{L-Platooning} quickly completes the vehicle join maneuver.}
\label{fig:join-time}
\end{figure}

We measured the time it takes for \emph{L-Platooning} to complete the vehicle join maneuver, \emph{i.e.,} from the point when the request message is sent by the joining vehicle and to the point when the desired inter-vehicle distance is achieved. For this experiment, we varied the distance to the front vehicle when the request message is sent. Results are depicted in Fig.~\ref{fig:join-time}. The result indicates that it takes slightly longer to complete the join maneuver when the request message is sent early, \emph{i.e.,} when the distance to the front vehicle is larger because the joining vehicle needs more time to reduce the inter-vehicle distance to 20m. Overall, we observe that \emph{L-Platooning} completes the vehicle join maneuver quickly with the average delay of 38sec in our simulations.

\subsection{Vehicle Leave Maneuver}
\label{sec:vehicle_leaving_performance}

In this section, we evaluate the performance of \emph{L-Platooning} focusing on how it handles the vehicle leave maneuver. We consider a scenario in which a randomly selected vehicle leaves from a long platoon. Specifically, the leaving vehicle sends a leave request to the platoon leader (or a virtual leader), and once it is confirmed by the platoon leader (or a virtual leader), it changes the lane and leaves from the platoon. In particular, if the leaving vehicle is a virtual leader, it also sends a message to the immediately following vehicle so that the following vehicle will serve as the virtual leader.

\begin{figure*}
\centering
\includegraphics[width=.99\textwidth]{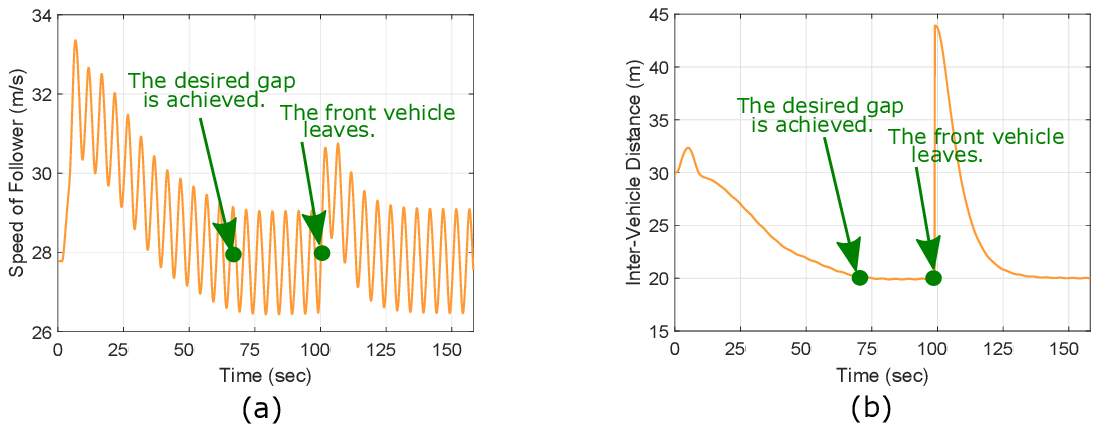}
\caption {(a) The speed of the immediately following vehicle. At 100sec, the speed is increased to reduce the gap created by the leaving vehicle; (b) The inter-vehicle distance of the immediately following vehicle. The inter-vehicle distance increases sharply as the vehicle leaves but is quickly adjusted to back to 20m.}
\label{fig:leave-speed_combined}
\end{figure*}


To validate that \emph{L-Platooning} successfully handles the vehicle leave maneuver for a long vehicle, we recorded the speed and inter-vehicle distance of the immediately following vehicle of the leaving vehicle. Fig.~\ref{fig:leave-speed_combined}(a) shows the speed of the following vehicle. It is shown that the vehicle increases the speed significantly to catch up with the platoon leader and to form a platoon. And when it is close enough to the platoon leader, it starts to adjust its speed according to that of the platoon leader which changes in a sinusoidal fashion. It is also observed that the speed of the vehicle gradually decreases so that the inter-vehicle distance slowly decreases down to the desired 20m. Note that the desired inter-vehicle distance is achieved at around 60sec. And then, the front vehicle leaves at 100sec. As the front vehicle leaves, the following vehicle quickly increases the speed to reduce the wider inter-vehicle gap created by the leaving vehicle and adjusts the inter-vehicle distance back to 20m.

Fig.~\ref{fig:leave-speed_combined}(b) shows the inter-vehicle distance of the following vehicle. The graph displays that the inter-vehicle distance keeps decreasing, and the desired inter-vehicle distance of 20m is achieved at around 60sec. When the front vehicle leaves at 100sec, it is observed that the inter-vehicle distance is sharply increased because of the gap created by the leaving vehicle. We also observe that \emph{L-Platooning} allows the vehicle to quickly adjust the large gap back to 20m.

Finally we measured the time for \emph{L-Platooning} to complete the vehicle leave maneuver. More specifically, a duration of time between the point when the leave request is sent to the platoon leader and the point when the immediately following vehicle finishes adjusting the inter-vehicle distance to 20m is measured. We repeated simulations 10 times for each randomly selected leaving vehicle to measure the delay. The results indicate that \emph{L-Platooning} is capable of handling the vehicle leave maneuver effectively and quickly with the average delay being 35.7sec with the standard deviation of 0.2sec.

\section{Discussion}
\label{sec:discussion}

In this section, we discuss other important performance issues of driving safety, driving comfort, and fuel efficiency specifically for managing and operating a long platoon. We note that each performance criteria may require different approaches compared to traditional platooning algorithms due to the unique characteristics of a long platoon.

\subsection{Driving Safety}
\label{sec:driving_safety}

A long platoon saves more fuel since more vehicles have reduced aerodynamic drag. However, the driving safety can be impacted because more vehicles travel with a very small inter-vehicle distance~{\cite{axelsson2016safety}}. The most obvious hazard is the possibility of a platoon member crashing into a preceding vehicle~{\cite{zheng2014study}}. In particular, a long platoon poses unique challenges of driving safety in terms of this type of hazard that opens the door for new research. For example, an existing solution based on a broad field-view sensor~{\cite{nowakowski2015cooperative}} may not be able to capture the platoon effectively due to the large size. A V2X-based solution~{\cite{alam2015heavy}} should be redesigned to deliver a safety message reliably to all platoon members of a long platoon within a very short time period.

Another important hazard for platooning is possible accidents due to vehicle's cut-in maneuver~{\cite{zheng2013safety}}. This type of hazard is significantly more dangerous for a long platoon because a long platoon takes a significant portion of a lane, making other vehicles very difficult to change lanes, although a long platoon is expected be used more frequently for nighttime freight logistics activities where there are few vehicles around for better efficiency. This safety challenge warrants new research directions. For example, an interactive approach can be developed that allows real-time communication with a long platoon and surrounding vehicles, so that the formation of a long platoon is dynamically reconfigured to allow vehicles to change lanes easily.

A sudden lane change of a platoon leader, {\emph{e.g.,}} to avoid an obstacle on a highway, can be very dangerous to following platoon members. The platoon leader has to disseminate such dangerous situation to the following vehicles in a very short time period, which can be especially challenging for a long platoon consisting of many platoon members where some of the platoon members are out of the communication range of the platoon leader.

\subsection{Driving Comfort}
\label{sec:driving_comfort}

While it is still debatable on how to define driving comfort, especially measuring the driving comport for automated vehicles, three most important factors in the literature are lane changes, acceleration, deceleration~{\cite{bellem2018comfort}}. Since lane changes occur only infrequently for platooning, the key factors influencing driving safety can be the degree of acceleration and deceleration~{\cite{marsden2001towards}}{\cite{deng2016general}}. According to a survey, it is known that the maximum acceleration and deceleration within the range of driving comfort are 2 {m/s$^2$} and 3 {m/s$^2$}, respectively~{\cite{hoberock1976survey}}. Given the simulation setting in Section~{\ref{sec:simulation_setup}}, we observe that the acceleration of all platoon members equipped with \emph{L-Platooning} were kept within the range of [-1.65, 1.57]m/s$^2$ despite the continuous sinusoidal speed change of the platoon leader.

Driving comfort for \emph{L-Platooning} is achieved by setting the maximum acceleration and deceleration for the CACC controller. However, we believe that better driving comfort can be achieved by making modifications at the protocol level, which is our future work. More specifically, in this paper, the quality of connectivity is a major factor that is used to select a virtual leader. A new metric representing driving comfort can be integrated into the model for selecting a virtual leader, being motivated by a latest research that the driving comfort differs depending on the distance from the platoon leader~{\cite{ma2020distributed}}, \emph{i.e.,} the variation of acceleration depends on the distance from the platoon leader.

\subsection{Fuel Efficacy}
\label{sec:traveling_efficacy}

\begin{figure}
\centering
\includegraphics[width=.9\columnwidth]{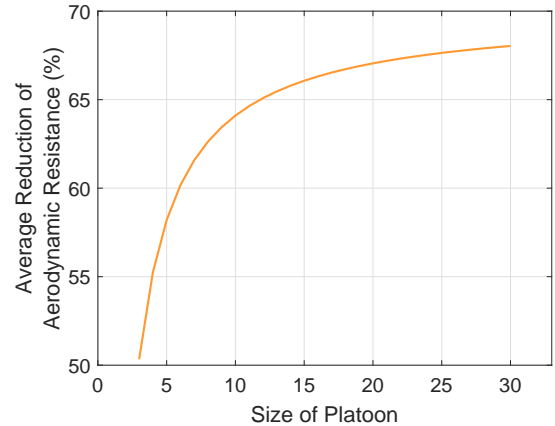}
\caption {Average air drag reduction for platoons with different sizes.}
\label{fig:fuel_efficiency}
\end{figure}

The fuel efficiency of platooning especially for heavyduty trucks has been investigated extensively~{\cite{lyamin2016study}}. Basically, platooning creates the split-stream effect for platoon members thereby decreasing the air-drag for the platoon members. Decreased air-drag, accounting for more than 23\% of the total forces imposed to a platoon member at a highway speed~{\cite{sandberg2001heavy}}, reduces the restrictive forces, consequently improving the fuel efficiency~{\cite{lyamin2016study}}. It is demonstrated that the air drag depends on the vehicle type, inter-vehicle distance, and the position in the platoon~{\cite{zabat1995aerodynamic}\cite{deng2016general}}. In particular, a smaller inter-vehicle distance leads to significantly improved fuel efficiency. Regarding the position in the platoon, the air drag for the first three to four vehicles is the highest, and the following vehicles have nearly the same small air drag~{\cite{hucho1993aerodynamics}}, indicating that creating a long platoon is significantly more fuel efficient as the following vehicles after the third or fourth platoon member will have small air drag. Specifically, our numerical analysis following Hucho \emph{et al.}~{\cite{hucho1993aerodynamics}} shows that the air drag reduction for a platoon with 30 heavy-duty trucks was 35.2\% higher than the one with 3 heavy-duty trucks as shown in Fig.~{\ref{fig:fuel_efficiency}}.

\section{Conclusion}
\label{sec:conclusion}

We presented \emph{L-Platooing}, the first protocol that enables seamless, reliable, and rapid formation of a long platoon, effectively addressing the current problem of the limited range of V2V communication for platooning. \emph{L-Platooning} allows platoon members maintain precisely the desired inter-vehicle distance regardless of the size of the platoon and effectively handles both the vehicle join and leave maneuvers. As the first protocol specifically designed to support long platooning, we expect that this work will be significant assets to the research community and industry especially for logistics company that have vast interests in deploying a platoon of large trailer trucks.

\bibliographystyle{IEEEtran}
\bibliography{mybibfile}

\begin{IEEEbiography}[{\includegraphics[width=1in,height=1.25in,clip,keepaspectratio]{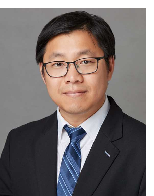}}]{Myounggyu Won}
(M'13) received a Ph.D. degree in Computer Science from Texas A\&M University at College Station, in 2013. He is an Assistant Professor in the Department of Computer Science at the University of Memphis, Memphis, TN, United States. Prior to joining the University of Memphis, he was an Assistant Professor in the Department of Electrical Engineering and Computer Science at the South Dakota State University, Brookings, SD, United States from Aug. 2015 to Aug. 2018, and he was a postdoctoral researcher in the Department of Information and Communication Engineering at Daegu Gyeongbuk Institute of Science and Technology (DGIST), South Korea from July 2013 to July 2014.  His research interests include smart sensor systems, connected vehicles, mobile computing, wireless sensor networks, and intelligent transportation systems. He received the Graduate Research Excellence Award from the Department of Computer Science and Engineering at Texas A\&M University - College Station in 2012.
\end{IEEEbiography}



\end{document}